\documentclass[twocolumn,showpacs,preprintnumbers,amsmath,amssymb,aps,floatfix,eps,superscriptaddress]{revtex4}

\usepackage{graphics}	
\usepackage{bm}		
\usepackage{alltt}	

\begin{document}

\title{Measuring storage and loss moduli using optical tweezers: broadband microrheology}

\author{Manlio Tassieri\footnote[2]{Electronic address: \texttt{M.Tassieri@elec.gla.ac.uk}}}
\affiliation{Department of Electronics and Electrical Engineering, University of Glasgow, G12 8LT, U.K.}

\author{Graham M. Gibson}
\affiliation{Department of Physics and Astronomy, SUPA, University of Glasgow, Glasgow, G12 8QQ, U.K.}

\author{R. M. L. Evans}
\affiliation{School of Physics and Astronomy, University of Leeds, Leeds, LS2 9JT, U.K.}

\author{Alison~M.~Yao}
\affiliation{Department of Physics and Astronomy, SUPA, University of Glasgow, Glasgow, G12 8QQ, U.K.}

\author{Rebecca Warren}
\affiliation{Department of Electronics and Electrical Engineering, University of Glasgow, G12 8LT, U.K.}

\author{Miles J. Padgett}
\affiliation{Department of Physics and Astronomy, SUPA, University of Glasgow, Glasgow, G12 8QQ, U.K.}

\author{Jonathan M. Cooper}
\affiliation{Department of Electronics and Electrical Engineering, University of Glasgow, G12 8LT, U.K.}

\date{15 July, 2009}

\begin{abstract}
We present an experimental procedure to perform broadband microrheological measurements with optical tweezers.~A generalised Langevin equation is adopted to relate the time-dependent trajectory of a particle in an imposed flow to the frequency-dependent moduli of the complex fluid. This procedure allows us to measure the material linear viscoelastic properties across the \emph{widest} frequency range achievable with optical tweezers.
\end{abstract}

\pacs{83.60.Bc, 66.20.-d, 83.50.-v, 83.85.Ei}

\maketitle

In 1986 Ashkin and colleagues reported the first observation of what is now commonly referred to as optical tweezers: a tightly focused beam of light capable of holding microscopic particles stable in three dimensions \cite{Ashkin86}.
Since then, several studies have adopted this technique as a tool for purposes as varied as trapping solid aerosols \cite{Summers}, measuring the viscosity of biomaterials \cite{Negulescu,Svoboda}, the forces exerted by single motor proteins \cite{Finer} and  the compliance of bacterial tails \cite{Block}, or stretching single DNA molecules \cite{Wang}. However, there remain a number of issues when optical tweezers are used for microrheological measurements.

Microrheology is a branch of rheology having the same principles as conventional bulk rheology (i.e. to study the linear viscoelastic behaviour of materials), but working on micron length scales. The linear viscoelastic properties of a material can be represented by the frequency-dependent dynamic complex modulus $G^*(\omega)$, which provides information on both the viscous and the elastic nature of the material. The conventional method of measuring $G^*(\omega)$ is based on the imposition of an oscillatory stress $\sigma(\omega,t)$ and the measurement of the resulting oscillatory strain $\gamma(\omega,t)$, \textit{or vice versa}. The amplitudes of its in-phase and out-of-phase components are both proportional to the stress amplitude, with constants of proportionality defining, respectively, the storage (elastic) $G'(\omega)$ and the loss (viscous) $G''(\omega)$ moduli \cite{Ferry}.

Optical tweezers have been successfully used with Newtonian fluids for rheological purposes such as determining the fluid viscosity with high accuracy, measuring the hydrodynamic interactions between particles or estimating the wall effect on the Stokes drag coefficient (i.e. Faxén's correction), as reviewed in Ref.~\cite{Yao}. Conversely, when optical tweezers are adopted for measuring the viscoelastic properties of complex fluids the results are limited to the material high frequency response, discarding the essential information related to long times scales (i.e. low frequency) material behaviour.

The aim of this letter is to present a self-consistent procedure for measuring the linear viscoelastic properties of materials, from non-oscillatory measurements, across the widest frequency range achievable with optical tweezers. In particular, the procedure consists of two steps: (I) measuring the thermal fluctuations of a trapped bead for a sufficiently long time; (II) measuring the transient bead displacement, from the optical trap centre, in response to a uniform fluid flow field entraining the bead. The flow is instantaneously switched on at time zero, by translating the whole fluid sample while the trap is held fixed. The imposed constant-velocity motion continues until a steady displacement of the bead is reached.
The analysis of the first step (I) provides: (a) the trap stiffness ($\kappa$) - note that this has the added advantage of making the present method self-calibrated - and (b) the high frequency viscoelastic properties of the material, to high accuracy. 
The second step (II) has the potential to provide information about the viscoelastic properties of the material down to very low frequencies, limited only by the duration of the experiment. However, because of the harmonic nature of the optical trap, that tends not to trasmit high-frequency applied forces to the bead, the material's high-frequency response can not be determined by this step. The full material viscoelastic spectrum is thus resolved by combining the results obtained from steps (I) and (II).

The new experimental procedure is analytically described through the analysis of the motion of a bead trapped in a stationary harmonic potential of force-constant $\kappa$, where a uniform fluid flow field of magnitude $V_s$ can be exerted at time $t=0$. The equation describing the bead position $\vec{r}(t)\;\forall\;t$ can be derived by means of the generalized Langevin equation, which in three dimensions is:
\begin{equation}
\label{Langevin}
	m \vec{a}(t)=\vec{f}_{R}(t)-\int^{t}_{0}\zeta(t-\tau)({\vec{v}}(\tau)-\vec{V}_s(\tau))d\tau-\kappa \vec{r}(t),
\end{equation}
where $m$ is the mass of the particle, $\vec{a}(t)$ is its acceleration, $\vec{v}(t)$ is the bead velocity, $\vec{V}_s(t)$ is the fluid flow field velocity and $\vec{f}_{R}(t)$ is the resultant of the stochastic thermal forces acting on the particle. The integral term represents the viscous damping of the fluid, which incorporates a generalized time-dependent memory function $\zeta(t)$.

We now show how Eq.~(\ref{Langevin}) evolves in the two cases mentioned above: when $\vec{V}_s(t)=0$ and $\vec{V}_s(t)\neq0$, respectively.

In the first case, where $\vec{V}_s(t)=0$, the optical tweezers can be calibrated by using the Principle of Equipartition of Energy:
\begin{equation}
\label{Equipartition}
	\frac{3}{2}k_{B}T=\frac{1}{2}\kappa \langle r^2 \rangle,
\end{equation}
where, $k_{B}$ is Boltzmann's constant, $T$ is absolute temperature and $\langle r^2 \rangle$ is the time-independent variance of the particle displacement from the trap centre, the origin of $\vec{r}$. Despite all the possible methods for determining the optical trap stiffness (e.g. using the power spectrum or the drag force \cite{Berg,Neuman,Tolic}), the Equipartition method is the only method independent of the viscoelastic properties of the material under investigation and is thus essential from a rheological point of view.

The thermal fluctuations of the trapped bead can also be investigated to determine the high frequency viscoelastic properties of the material through analysis of the time dependence of the mean-square displacement $\left\langle \Delta r^{2}(\tau)\right\rangle$ (MSD):
\begin{equation}
\label{MSD}
	\left\langle \Delta r^{2}(\tau)\right\rangle \equiv 
	\left\langle\left[\vec{r}(t+\tau)-\vec{r}(t)\right]^{2}\right\rangle_{t},
\end{equation}
where $t$ is absolute time and $\tau$ is the lag-time. The average is taken over all initial times $t$ and the number of particles considered in the experiment, if more than one. In particular, using the assumptions adopted by Mason and Weitz in the study of the motion of thermally excited free particles \cite{Mason}, at thermal equilibrium, where $\left\langle {\vec{v}}(t)\vec{f}_{R}(t)\right\rangle=0$ and $m \left\langle {\vec{v}}(t){\vec{v}}(t)\right\rangle=6k_{B}T\;\forall\;t$, Equation (\ref{Langevin}) yields, in the Laplace form, the velocity autocorrelation function:
\begin{equation}
\label{LL}
	\left\langle {v}(0){\tilde{v}}(s)\right\rangle=\frac{6k_{B}T}{ms+\tilde{\zeta}(s)+\kappa/s}\equiv s^{2}\left\langle \Delta\tilde{r}^{2}(s)\right\rangle,
\end{equation}
where $s$ is the Laplace frequency. Following Mason and Weitz \cite{Mason} in assuming that the bulk Laplace-frequency-dependent viscosity of the fluid $\tilde{\eta}(s)$ is proportional to the microscopic memory function $\tilde{\zeta}(s)=6\pi a \tilde{\eta}(s)$, where $a$ is the bead radius, Eq. (\ref{LL}) can be written as:
\begin{equation}
\label{CM}
	\tilde{\eta}(s)=\frac{1}{6\pi a}\left[\frac{6k_{B}T}{s^{2}\left\langle \Delta\tilde{r}^{2}(s)\right\rangle}-ms-\frac{\kappa}{s}\right],
\end{equation}
where the first term in the brackets reflects the viscoelasticity of the medium, the second term is related to the inertia of the bead and the third term takes into account the optical trap strength. It is easy to demonstrate that, for a micro-bead of density of order of $1 g/cm^3$ suspended in water, the product $ms$ is negligible compared with the first term for the majority of the experimentally accessible frequencies (i.e. $s << 10^{6}$ s$^{-1}$). With regard to the optical trap strength, two limiting cases can be distinguished: (i) in the limit $\kappa/s \rightarrow 0$, which can be obtained either for vanishing trap strength or for measurements performed at high frequencies, but lower than $10^{6}$ s$^{-1}$, Eq. (\ref{CM}) recovers the generalized Stokes-Einstein relationship derived by Mason and Weitz \cite{Mason}; (ii) in the limit $\kappa/s \rightarrow \infty$, which can be obtained either for a strong optical trap or for measurements performed at very low frequencies, Eq. (\ref{CM}) gives the same result as if the bead were embedded in a purely elastic continuum with elastic constant of $\kappa/6\pi a$. For all intermediate cases, where $0 <\kappa/s< \infty$, it is easy to show that, by analytic continuation from Eq.~(\ref{CM}), the complex modulus can be expressed directly in terms of the time-dependent MSD:
\begin{equation}
\label{G*MSD}
	G^*(\omega) = \left. s\tilde{\eta}(s) \right|_{s=i\omega}=\frac{\kappa}{6\pi a}\left[\frac{2\langle r^2\rangle }{i\omega \left\langle\Delta \widehat{r^2}(\omega)\right\rangle}-1 \right],
\end{equation}
where $\left\langle\Delta \widehat{r^2}(\omega)\right\rangle$ is the Fourier transform of $\langle\Delta r^2(\tau)\rangle$.

The second step of the procedure, which experimentally follows the first, consists of the analysis of the induced bead displacement from the trap centre due to an imposed time-dependent uniform fluid flow field $\vec{V}_s(t)$ entraining the bead.
In this case, Equation (\ref{Langevin}) yields, in the Laplace form, the mean velocity of the particle:
\begin{equation}
\label{mv}
	\left\langle {\tilde{v}}(s)\right\rangle=\frac{\tilde{\zeta}(s)\tilde{V}_s(s)}{ms+\tilde{\zeta}(s)+\kappa/s}\equiv s\left\langle \tilde{r}(s)\right\rangle,
\end{equation}
where the brackets $\left\langle ...\right\rangle$ denotes the average over several independent measurements (but not averaged over absolute time, since time-translation invariance has been broken by the flow start-up at $t=0$). 
It is straight-forward to show that, by analytic continuation from Eq.~(\ref{mv}), the complex modulus can be expressed directly in terms of both the imposed flow field and of the induced bead displacement from the trap centre:
\begin{equation}
\label{G*r}
	G^*(\omega) = \left. s\tilde{\eta}(s) \right|_{s=i\omega}=\frac{(\kappa-m\omega^2)i\omega\left\langle \hat{r}(\omega)\right\rangle}{6\pi a\left(\hat{V}_s(\omega)-i\omega\left\langle \hat{r}(\omega)\right\rangle\right)},
\end{equation}
where $\hat{V}_s(\omega)$ and $\left\langle \hat{r}(\omega)\right\rangle$ are the Fourier transforms of $\vec{V}_s(t)$ and $\left\langle \vec{r}(t)\right\rangle$, respectively.
Note that, so far, the temporal form of $\vec{V}_s(t)$ is still undefined. Thus Eq.~(\ref{G*r}) represents the general solution for $G^*(\omega)$ independently of the temporal form of $\vec{V}_s(t)$ (e.g. sinusoidal function $\vec{V}_s \sin(\omega t)$ or, as in this work, Heaviside step function $\vec{V}_s H(t)$, where $\vec{V}_s(t)=0\;\forall\;t<0$ and $\vec{V}_s(t)= \vec{V}_s \;\forall\;t\geq0$).

In principle, Equations~(\ref{G*MSD}) and (\ref{G*r}) are two simple expressions relating the material complex modulus $G^*(\omega)$ to the observed time-dependent bead trajectory $\vec{r}(t)$ \textit{via} the Fourier transform of either the $\vec{r}(t)$ itself (in Eq. (\ref{G*r})) or the related MSD (in Eq. (\ref{G*MSD})).
In practice, the evaluation of these Fourier transforms, given only a finite set of data points over a finite time domain, is non-trivial since interpolation and extrapolation from those data can yield serious artifacts if handled carelessly.

In order to express the two Fourier transforms in Equations (\ref{G*MSD}) and (\ref{G*r}) in terms of the $N$ experimental data points $(t_k,\langle\Delta r^2(\tau)\rangle_k)$ and $(t_k,\left\langle \vec{r}(t)\right\rangle_k)$, respectively, where \mbox{$k=1\ldots N$}, which extend over a finite range, exist only for positive $t$ and {\em need not} be equally spaced, we adopt the analytical method introduced in Ref.~\cite{Evans}. In particular, we refer to Eq.~(10) of Ref.~\cite{Evans} which is equally applicable to find the Fourier transform $\hat{g}(\omega)$ of any time-dependent quantity $g(t)$ sampled at a finite set of data points $(t_k,g_k)$, giving:
\begin{eqnarray}
\label{Fg}
  -\omega^2 \hat{g}\left(\omega\right) =  i\omega g(0) + \left( 1-e^{-i\omega t_1}\right) \frac{\left(g_1-g(0)\right)}{t_1} \nonumber \\
  + \dot{g}_\infty e^{-i\omega t_N} + \sum_{k=2}^N \left( \frac{g_k-g_{k-1}}{t_k-t_{k-1}} \right) \left( e^{-i\omega t_{k-1}}-e^{-i\omega t_k} \right),
\end{eqnarray}
where $\dot{g }_\infty$ is the gradient of $g (t)$ extrapolated to infinite time. Also $g (0)$ is the value of $g(t)$ extrapolated to $t=0^+$.
Identical formulas can be written for both $\left\langle\Delta \widehat{r^2}(\omega)\right\rangle$ and $\left\langle \hat{r}(\omega)\right\rangle$, with $g$ replaced by $\langle\Delta r^2\rangle$ and $\left\langle \vec{r}\right\rangle$, respectively. This analytical procedure has the advantage of removing the need for Laplace/inverse-Laplace transformations of experimental data \cite{Mason2}.

We have tested Equations (\ref{G*MSD}) and (\ref{G*r}), \textit{via} Eq. (\ref{Fg}),  by measuring both the viscosity of water and the viscoelastic properties of water-based solutions of polyacrylamide (PAM, flexible polyelectrolytes, $M_w=5$~---~$6\times10^6$~g/mol, Polysciences Inc.)  using optical tweezers as described below.

Trapping is achieved using a CW Ti:sapphire laser system (M Squared, SolsTiS) which provides up to 1Watt at 830nm. The tweezers are based around an inverted microscope, where the same objective lens, 100$\times$ 1.3NA, (Zeiss, Plan-Neofluor) is used both to focus the trapping beam and to image the resulting motion of the particles. Samples are mounted in a motorized microscope stage (ASI, MS-2000). Two CMOS cameras are used to view the sample, with bright-field illumination; one provides a wide field of view (Prosilica EC1280M), while the other takes high speed images of a reduced field of view (Prosilica GV640M). These images are processed in real time at 2kHz using our own LabVIEW (National Instruments) particle tracking software running on a standard desktop PC~\cite{Gibson}.

The Brownian fluctuations of an optically trapped bead give rise to the time dependent $\langle\Delta r^2(\tau)\rangle$ shown in the inset of Fig.~\ref{MSD}. In the case of a bead immersed in a Newtonian fluid, it is expected that at short time intervals (thus small distances) the bead behaves as if it were free to diffuse. Indeed, the agreement between the observed $\langle\Delta r^2(\tau)\rangle$ at short times of a trapped bead in water (circles) and the Einstein prediction for a freely diffusing bead (solid line) is good. As the time intervals increase the bead becomes influenced by the optical potential. This results in a plateau  at large time intervals, where the $\langle\Delta r^2(\tau)\rangle$ tends to $2\langle r^2\rangle$. It is interesting to note that the ratio of these two quantities (the MSD and twice the variance of the positional distribution) is a dimensionless parameter, independent of both the optical trap stiffness and the bead radius. It thus allows an explicit comparison between the dynamics of the fluids under investigation (Fig.~\ref{MSD}). Moreover, the onset point of the plateau region in Fig.~\ref{MSD} indicates the bottom limit of the frequency range within which the moduli can be determined by Eq.~(\ref{G*MSD}), as for all the previous works using stationary optical tweezers.
\begin{figure}[t]
	\centering
		\resizebox{80mm}{!}{\includegraphics{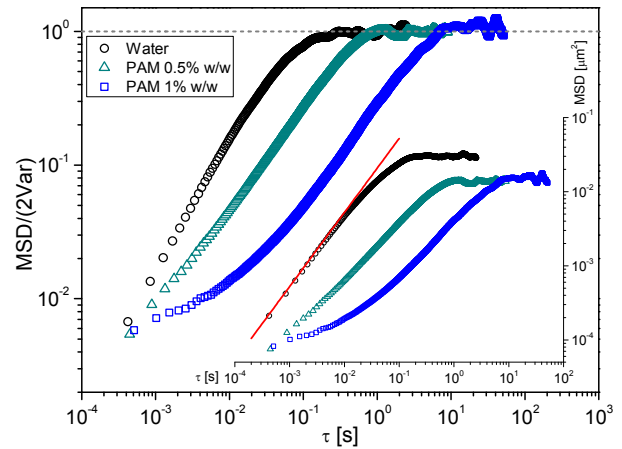}}
	\caption{\label{MSD}(Color online) The normalised MSD \textit{vs.} lag-time of a $5 \mu m$ diameter bead in water and in two water based solutions of PAM at concentrations of $0.5$ \% $w/w$ and $1$ \% $w/w$. The inset shows the same data, unnormalised. The line in the inset is the Einstein prediction of the MSD for a $5 \mu m$ diameter bead in water at $25^o C$.}
\end{figure}

\begin{figure}[h]
	\centering
		\resizebox{80mm}{!}{\includegraphics{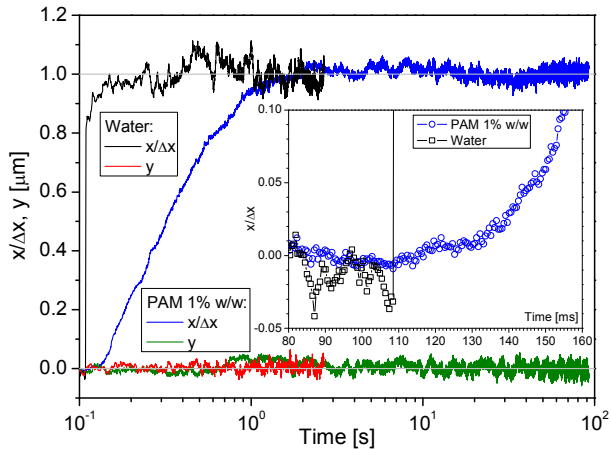}}
	\caption{\label{Step}(Color online) The coordinates of a $5 \mu m$ diameter bead \textit{vs.} time for two different solutions and for two uniform fluid flow fields of different magnitude $V_s$ at $25^o C$. In both the cases the data were averaged over three measurements and the $x$ coordinate has been normalised by the steady state displacement $\Delta x$. In water: $V_s = 20 \mu m/s$ and $\Delta x=0.523 \mu m$. In 1\% w/w of PAM, $V_s =3 \mu m/s$ and $\Delta x=1.155 \mu m$. The inset highlights the start-up behaviour of both the above systems.}
\end{figure}

In Figure~\ref{Step} we compare the responses of a $5 \mu m$ diameter bead immersed in water (a Newtonian fluid) and in a water solution of PAM at 1\% w/w (a non-Newtonian fluid), due to the imposition of a uniform fluid flow field having temporal behaviour as a Heaviside step function $\vec{V}_s(t)=\vec{V}_s H(t)$, with different magnitude in the two measurements. Experimentally, the execution of a Heaviside step function is achieved by suddenly moving the motorised microscope stage at a predetermined speed and direction (here parallel to the $x$ axis). The experiment runs until a steady displacement ($\Delta x$) of the bead from the trap centre is reached (i.e.~until all the material's characteristic relaxation times are exceeded). In Figure~\ref{Step} the $x$ component of the bead displacement has been normalised by $\Delta x$ for a better comparison between the viscoelastic character of the two samples. It is clear that while the Newtonian fluid reaches a steady value of the displacement almost \emph{instantaneously} (as expected), the non-Newtonian fluid shows complex dynamics representative of its viscoelastic nature.
It is important at this point to note that, because of the harmonic nature of the optical potential, at early times (i.e. for $t\rightarrow0$ or equivalently for $\omega\rightarrow\infty$), the trapping force exerted on the bead is actually small (i.e. $\kappa \vec{r}(t)\rightarrow0$) and the particle moves almost at the same speed as the imposed flow (i.e. $\vec{v}(t)\cong\vec{V}_s(t)$); this implies that Eq.~(\ref{G*r}) becomes undefined at high frequencies.

The broadband microrheological measurement with optical tweezers is achieved by combining the frequency responses obtained from both the methods introduced above.
In particular, the material's high frequency response is determined by applying Eq.~(\ref{G*MSD}) (\textit{via} Eq.~(\ref{Fg}) with $\langle\Delta r^2\rangle_k$ replacing $g_k$) to the $\langle\Delta r^2(\tau)\rangle$ measurements; whereas, the low frequency response is resolved by applying Equation~(\ref{G*r}) (\textit{via} Eq.~(\ref{Fg}) with $\left\langle \vec{r}\right\rangle_k$ replacing $g_k$) to the data describing the bead's transient response to the motion of the stage.

\begin{figure}[h]
	\centering
		\resizebox{80mm}{!}{\includegraphics{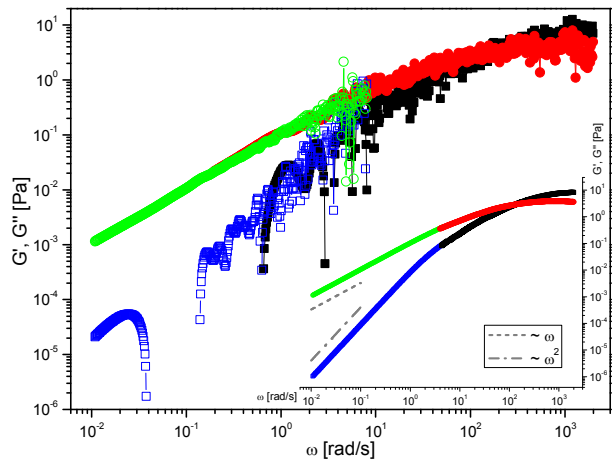}}
	\caption{\label{G'&G''}(Color online) Storage (squares) and loss (circles) moduli \textit{vs.} frequency of a solution of 1\% w/w of PAM in water measured by means of both Eq.~(\ref{G*MSD}) (solid symbols) and Eq.~(\ref{G*r}) (open symbols) applied directly to the experimental data presented in Fig. \ref{MSD} and Fig. \ref{Step}, respectively. The inset shows the moduli of the same solution as above, with both Eq.~(\ref{G*MSD}) and Eq.~(\ref{G*r}) applied to the original data mentioned above, but smoothed.}
\end{figure}

A typical result of this procedure for a non-Newtonian fluid is shown in Figure~\ref{G'&G''} while, in the case of water, a constant viscosity of $\eta = 8.69\times10^{-4} \pm 6\times 10^{-6} Pa\cdot s$ is measured over five frequency decades at $25^o C$. It is evident that, although there is some noise in the frequency domain that has propagated from genuine experimental noise in the time-domain data, there is a clear overlapping region of agreement between the two methods that makes the whole procedure self-consistent. Moreover, it confirms the ease with which the low-frequency material response can be explored, right down to the terminal region (where $G' \propto \omega^{2}$ and $G'' \propto \omega$), which is the current limitation for microrheological measurements performed with optical tweezers. However, in order to remove the genuine noise, a simple smoothing operation of the original data is sufficient and the results are shown in the inset of Figure~\ref{G'&G''}.

In summary, we have presented a self-consistent and simple experimental procedure, coupled with an analytical data analysis method, for determining the broadband viscoelastic properties of complex fluids with optical tweezers. This method extends the range of the frequency response achieved by conventional optical tweezers measurements down to the material's terminal region.

ACKNOWLEDGMENTS: The project was funded by the BBSRC and EPSRC DTC. RMLE is funded by the Royal Society.

\end{document}